\documentclass[twocolumn,showpacs,preprintnumbers,amsmath,amssymb,nofootinbib]{revtex4}

\usepackage{graphicx}
\usepackage{dcolumn}
\usepackage{bm}
\usepackage{epsf}

\begin{document}


\title{Revisiting algorithms for generating surrogate time series}

\author{C. R\"ath$^1$, M. Gliozzi$^2$, I. E. Papadakis$^{3,4}$, W. Brinkmann$^1$}
 \affiliation{ $^1$ Max-Planck Institut f\"ur extraterrestrische Physik, Giessenbachstr. 1, 85748 Garching, Germany\\
        $^2$ George Mason University, Department of Physics and Astronomy, MS 3F3, 4400 University Dr., Fairfax, VA 22030-4444, USA \\
        $^3$ Physics Department, University of Crete, PO Box 2208, 710 03 Heraklion, Crete, Greece \\
        $^4$ IESL, Foundation for Research and Technology, 711 10, Heraklion, Crete, Greece
        }

\date{\today}

\begin{abstract}
The method of surrogates is one of the key concepts of nonlinear data analysis.
Here, we demonstrate that commonly used algorithms for generating surrogates
often fail to generate truly linear time series.
Rather, they create surrogate realizations with Fourier phase correlations
leading to non-detections of nonlinearities. We argue that reliable surrogates can only be
generated, if one tests separately for static and dynamic nonlinearities.

\end{abstract}



\maketitle

{\it Introduction.} The method of surrogates \cite{theiler92}  is one of the the key concepts
of nonlinear data analysis, which allows to test for weak nonlinearities in data sets in 
a model-independent way. 
The basic idea of this approach is to compute statistics sensitive 
to nonlinearities for the original data set and for an ensemble of so-called surrogate data sets,
which mimic the linear properties of the original data.
If the computed measure for the
original data is significantly different from the values obtained
for the set of surrogates, one can infer that the data contain nonlinearities.\\
Linearity means in this case that all the structure in the time series 
is contained in the autocorrelation function, or equivalently, in 
the Fourier power spectrum. Thus the time series $y_t$ can be modeled e.g. 
by an autoregressive (AR) model described by 
$y_t = a_0 + \sum_{k=1}^{q} a_{k} y_{t-k} +  \sigma e_t$ ($a$: coefficients, $e$: white noise) 
or a  more general ARMA models also including moving averages (MA). 
Nonlinearity thus refers to all those structures in data sets that are not captured 
by the power spectrum.\\
Since its introduction the method of surrogates has found numerous applications in many fields
of research ranging from geophysical and  physiological time series analysis  \cite{Schreiber00} 
to econophysics \cite{Wang08}, astrophysics \cite{Gliozzi10}, and cosmology \cite{Raeth09,Raeth11}.\\
The most commonly used methods for generating surrogates include 
Fourier-transformed (FT), amplitude adjusted Fourier-transformed (AAFT) \cite{theiler92} 
and iterative amplitude adjusted Fourier-transformed (IAAFT) surrogates \cite{schreiber96}.
With FT surrogates, which are generated by using the Fourier amplitudes of the original 
time series and a set of random phases, one tests whether the time series is compatible 
with a linear Gaussian process. In other words, any significant deviation of the original 
data from its surrogates indicates  the presence of higher order temporal correlations, 
i.e. the presence of dynamic nonlinearities.\\ 
The null hypothesis can be generalized  to cases where the data set is non-Gaussian. Here,
the hypothesis is that the original data $ \left\{  y_n \right\}  $ 
are from a nonlinear or linear stochastic process that has undergone
a static nonlinear transformation.  The AAFT algorithm was developed to generate this kind 
of surrogates. The original time series  is rendered Gaussian by a rank-ordered remapping of the 
values onto  a Gaussian distribution. For this Gaussian time series which follows
the measured time evolution, a FT surrogate time series is generated. The final 
surrogate data are obtained by rank-ordered remapping of the FT surrogate onto
the distribution of the original data.\\
It was shown that the final remapping step in the AAFT algorithm can lead to a
whitening of the power spectrum \cite{schreiber96}. 
This shortcoming was overcome by the IAAFT algorithm, which consists of the 
following iteration scheme. (1) One starts with a random shuffle  $\left\{ y_n^i \right\} $
of the original data $\left\{ y_n \right\} $,
which is Fourier transformed yielding the Fourier amplitudes $\left|  S^i(k)  \right| $ 
and Fourier phases $\phi^{i}(k)$ .  (2) The Fourier amplitudes $\left|  S^i(k)  \right| $ of the
time series $\left\{ y_n^i \right\} $ are then replaced by those for the 
original data  $\left| Y(k)  \right|  = \left|  \sum_{n=1}^{N} y_n e^{i2\pi kn/N}  \right| $. 
The phases of the complex Fourier  $\phi^{i}(k)$ components are kept.
An inverse Fourier transformation leads to a time series with desired power spectrum $\left\{ z_n^i \right\} $.
(3) To make the surrogate time series have the same amplitude distribution $\left\{y_n  \right\}$ in real space 
as the original data,   $\left\{ z_n^i \right\} $ is remapped in a rank ordered 
way onto  $\left\{ y_n^i \right\} $ leading to  $\left\{ y_n^{i+1} \right\} $.
As in AAFT the final remapping step changes the power spectrum, so that steps (2) and (3) have to be repeated
until convergence to the correct power spectrum is achieved.  
AAFT and IAAFT strive for  a proper reproduction of the 
linear properties and the amplitude distribution 
of the time series. 
All three algorithms have in common that they intend to translate the 
absence of nonlinear temporal correlations  into the constraint of uncorrelated and 
uniformly distributed phases in the Fourier representation of the data.
While the randomness of the phases remains (by construction) exactly 
preserved  for FT surrogates, it has never been studied, how the remapping step
(in AAFT) and the iteration steps (in IAAFT) affect the randomness of the phases
in the resulting surrogate time series.\\
In this {\it Letter} we investigate in detail how well this much more important 
constraint of the absence of nonlinear correlations in the surrogates is 
fulfilled for AAFT and IAAFT surrogates and study consequences for the
outcome of tests on nonlinearity for observational data.

\begin{figure}[h]
\centering
  \includegraphics[width=8.5cm,angle=0]{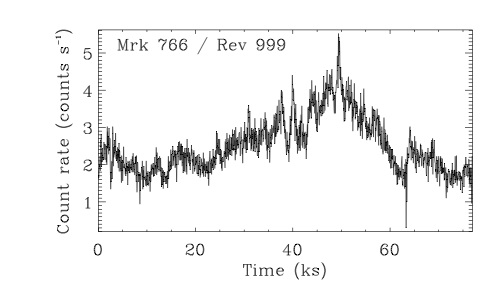}
  \includegraphics[width=8.5cm,angle=0]{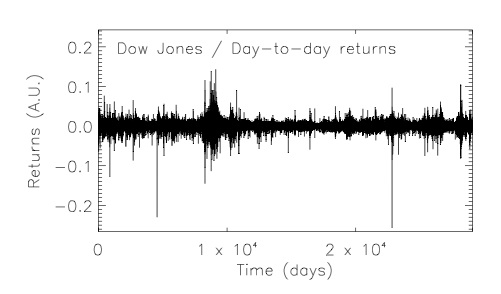}
  \caption{Light curve of the observation of Mrk 766 in 50 s time resolution (above) and the 
                  day-to-day returns  of the Dow Jones Industrial index (below). The time series consists of $1540$ and $29070$ points 
                  respectively.   \label{figure1}}
\end{figure}

{\it Data Set  and Preprocessing.} All following investigations are based 
on two -- quite distinct -- data sets. The first one is an X-ray observation of the 
narrow-line Seyfert 1 galaxy Mrk 766 taken with the XMM Newton satellite \cite{Markowitz07}. 
For light curves from active galactic nuclei (AGN) one investigates 
the linear and nonlinear temporal correlations to infer more information about the physical 
processes in the innermost regions of these compact accreting objects. 
The time series originates from observations of Mrk 766 in 2005. We selected the 
revolution 999 from May, 23rd, 2005 that took 95 ks. 
We used the data retrieved from the XMM public archive and 
only relied on data from the PN camera  \cite{Strueder01} 
due to its superior statistical quality.
Source counts were accumulated from a rectangular box of
27 $\times$ 26 RAW pixels  (1 RAW pixel $\sim$ 4.1 arcsec)
around  the  position of the source. Background data were extracted from a
similar,  source free region on the chip.
After rejection of times  affected by high background a light curve was extracted 
in the 0.3$-$10 keV energy band, background subtracted and binned to 50\,s time resolution.\\
The second data set are the day-to-day returns of the Dow Jones (DJ) industrial index covering the period from
26th of May 1896 until 8th of June 2012. From the publicly available (www.djaverages.com) 
closing prices $p(t)$ of each trading day
the day-to-day returns $r(t)$ are determined by considering the change of the
logarithm of the stock price $r_{\Delta t}(t) = \ln(p(t)) -\ln(p(t-\Delta t))$, where $\Delta t =1$ day. 
Both  data sets (see  Fig. \ref{figure1}) are particularly well suited to be used 
for testing concepts of nonlinear time series analysis, 
because they represent scalar, {\it real} time series from a sufficiently complex system, 
where the mere detection of  nonlinearities already allows to discriminate between different 
classes of models. In the case of the  the AGN-data comparably simple intrinsically linear models for light curve variations 
like e.g. global disk oscillations models \cite{Titarchuk00} can be safely ruled out.
The detection of signatures of nonlinear determinism in financial data shows that the 
stock market is neither completely governed by stochastic processes nor can it be 
purely described by linear processes plus additive noise.

\begin{figure}[h]
\centering
  \includegraphics[width=4.25cm,angle=0]{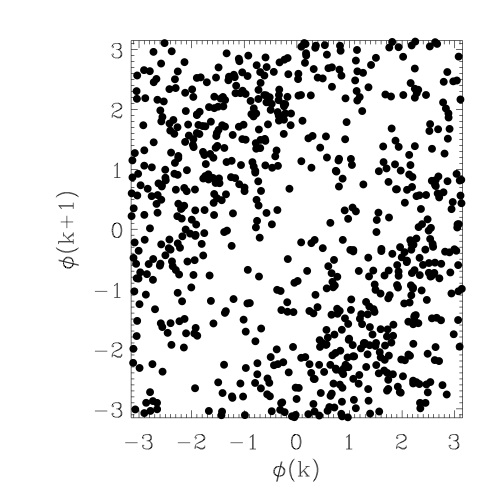}
  \includegraphics[width=4.25cm,angle=0]{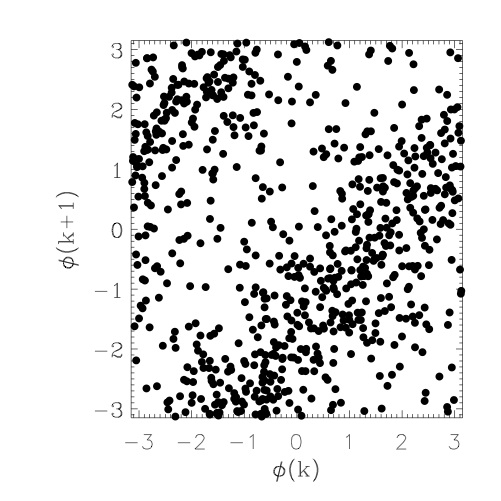}
   \includegraphics[width=4.25cm,angle=0]{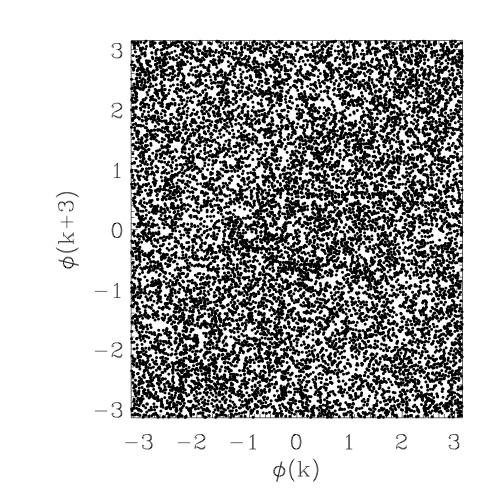}
  \includegraphics[width=4.25cm,angle=0]{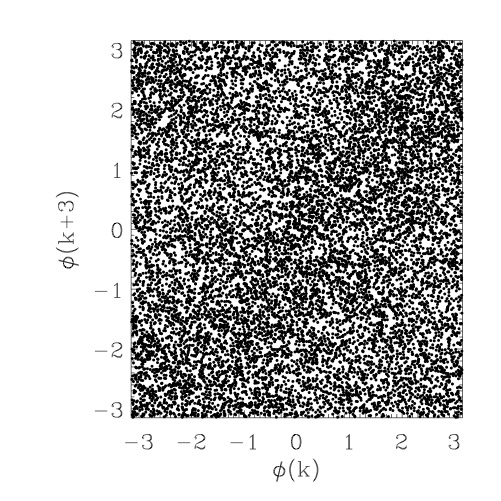}
  \caption{Phase maps  for $\Delta =1$ for one AAFT (left) and one IAAFT (right) surrogate realization for the 
  AGN time series (above). Below the phase maps for  $\Delta =3$ for one AAFT (left) and one 
  IAAFT (right) surrogate realization for the Dow Jones data are shown. \label{figure2}}
\end{figure}

\begin{figure}[h]
\centering
  \includegraphics[width=8.5cm,angle=0]{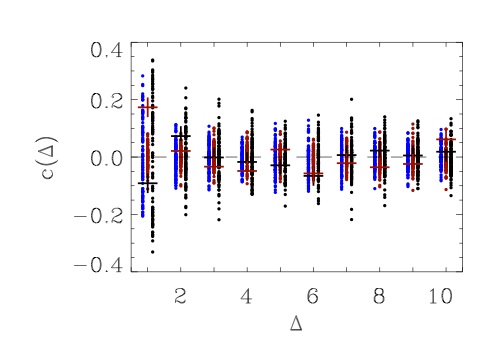}
   \includegraphics[width=8.5cm,angle=0]{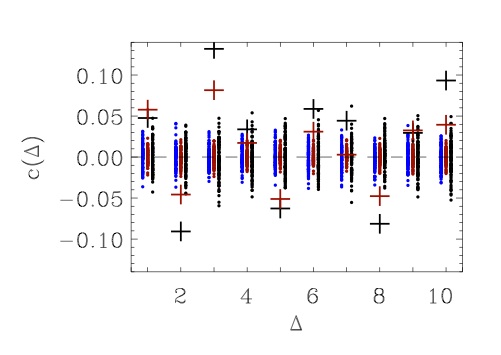}
  \caption{Cross-correlation between $\phi(k)$ and $\phi(k+\Delta)$ for  $\Delta =1,\ldots,10$ for 100 
  realizations of FT (red), AAFT (blue) and IAAFT (black) surrogates for the Mrk 766 (above) and the Dow Jones (below)
  time series. The black (red) crosses denote the values 
  for $c(\Delta)$ for the original time series (remapped original time series) for comparison.  \label{figure3}}
\end{figure}

\begin{figure}[h]
\centering
    \includegraphics[width=4.25cm,angle=0]{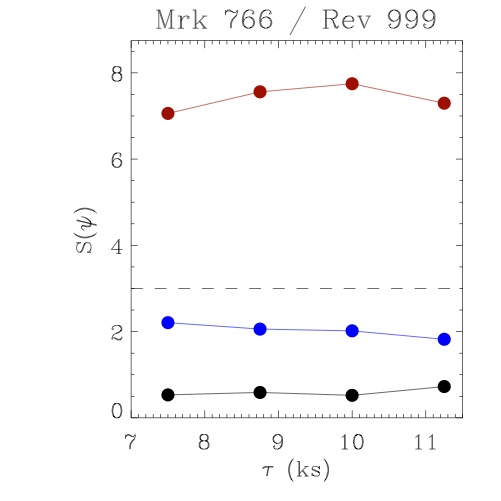}
     \includegraphics[width=4.25cm,angle=0]{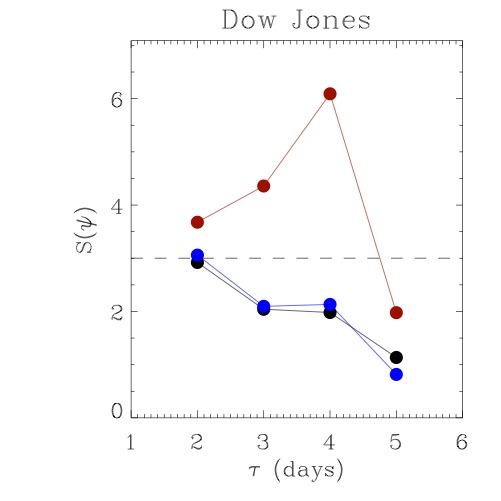}
  \caption{Significances $S(\psi)$ for nonlinear temporal correlations for the set of light curves as 
  determined with NLPE and as a function of the delay time $\tau$ for Mrk 766 (left) and the Dow Jones (right).
  Red: results obtained using the remapped time series and FT surrogates. 
  Black: results obtained with the original data and the ITAAFT 
  algorithm. Blue: results for the original data and the AAFT algorithm. 
  The dashed line indicates the 3 $\sigma$  detection limit. \label{figure4}}
\end{figure}

{\it Analyses and Results.} 
For both  time series we generate 200 realizations of FT,  AAFT and IAAFT surrogates each. 
Note that in the FT case the random phase hypothesis for the surrogates is doubtlessly fulfilled.
To assess the presence or absence of nonlinear correlations in the surrogates we make use of the 
representation of Fourier phases in so-called phase maps  \cite{chiang02}. 
A  phase map is defined as a two-dimensional set of points $G=\{ \phi_k,\phi_{k+\Delta} \}$ where $\phi_k$ is the 
phase of the $k^{th}$ mode of the Fourier transform and $\Delta$ a phase delay. 
If the phases were taken from a  random, uniform and uncorrelated distribution, the phase maps $G$ would be a random 
two-dimensional distribution of points in the interval $\left[ -\pi; \pi \right]$. Any significant deviation
from this random distribution already points towards the presence of nonlinearities in the data.
In Fig. \ref{figure2} we show the phase maps for the AGN case for a delay $\Delta =1$ for one AAFT and one IAAFT
realization. For the Dow Jones data we show the respective plots for $\Delta =3$. 
IFor both types of times series and for both 
classes of surrogates the phase maps are far from being random.
To quantify the degree of correlation between the phases $\phi$ and $\phi + \Delta$ we 
calculate the correlation coefficient
$c(\Delta)$ for delays up to $\Delta = 10$. In Fig. \ref{figure3} we show $c(\Delta)$ for the three 
classes of surrogates and both time series under study.
The jitter around zero of the FT surrogates (red)
represents -- by construction -- the statistical fluctuations of $c(\Delta)$ 
for a limited number of phases. 
For both examples, however,  these fluctuations around zero are significantly larger for the AAFT 
surrogates (blue) and IAAFT surrogates (black)  for a number of values of  $\Delta$. 
This results in up to $138$ (IAAFT) and $110$ (AAFT) surrogate realizations leading 
to $c$-values lying outside the 3 $\sigma$ confidence region around zero. 
Since the only difference between FT and AAFT is the remapping step
after the inverse Fourier transformation, we can immediately infer that this step is responsible for inducing the phase 
correlations.  The deviations from zero for the correlation coefficient become in both cases more pronounced
for the IAAFT surrogates. For the AGN light curve
one can even observe that for  $\Delta = 1$ the region around zero for $c(\Delta)$ is thinned out meaning that
it is rather a rule than an exception that $\left | c(\Delta) \right |$ becomes larger than 
the normal statistical fluctuations.
The presence of phase correlations in the majority of surrogate realizations
proves already in a strict mathematical sense that those realizations are in fact
nonlinear.\\
To address the relationship between phase correlations and measures for nonlinearity
 we calculate the nonlinear prediction error (NPLE) as described in \cite{Sugihara90}. 
The NLPE has been shown to 
be a robust measure with a good overall performance \cite{Schreiber97}.  
The calculation of the NLPE 
relies on the representation of the time series in an artificial phase space, which is obtained
using the method of delay coordinates \cite{Packard80}. 
This is accomplished by using time delayed versions of the observed time series as
coordinates for the embedding space. The multivariate vectors in the $d$-dimensional 
space are expressed by $\vec{y}_n= (y_{n-(d-1) \tau}, y_{n-(d-2)\tau}, \ldots , y_n ) \;,$
where $\tau$ is the delay time and $y_n$ denotes the value  of the (discretized) time series at time step $n$.
The comparison of the predicted behavior of the time series based on the local neighbors
with the real trajectory of the system leads to the definition of the NLPE $\psi$ as
\begin{eqnarray}
\psi & = & \psi(d,\tau,T,N) \\ \nonumber
           &= &\frac{1}{(M-T-(d-1)\tau)} \sqrt{ \sum_{n=(d-1)\tau}^{M-1-T} [\vec{y}_{n+T} -F(\vec{y}_n)]^2}      
\end{eqnarray}
where $F$ is a locally constant predictor, $M$ is the length of the time series, 
and $T$ is the lead time. The predictor $F$ is calculated by averaging over future values 
of the $N \, (N=d+1)$ nearest neighbors in the delay coordinate representation.\\ 
We found that for $T > 5$ $\psi$ remains rather constant, 
thus a value of $T=5$ was used for this study.
The dimension of the embedding space $d$ 
and the delay time $\tau$ have to be set appropriately. 
Since the AGN time series of this study consist only of $1539$ data points , 
we use a low embedding dimension $d=3$, so that the
vectors  $\vec{y}_n$ are not too sparsely distributed in the embedding space. To allow for direct comparison
we also used $d=3$ for the much longer time series of the Dow Jones.\\
The delay time $\tau$ is determined by using the criterion of zero crossing of the 
autocorrelation function or the first minimum of the autocorrelation function  or 
mutual information \cite{Fraser86},
which lies around $\tau= 10$ ks for Mrk 766 and  $\tau= 2$ days respectively.
To also investigate the robustness of our results with respect to
variations of $\tau$, 
we calculate the NPLE for a set of four different 
delay-times, $\tau= 7.5, 8.75, 10.0$ and $11.25$ ks (AGN data)
and  $\tau= 2, 3, 4$ and $5$ days (DJ data)
and analyze the results as a function of $\tau$.\\ 
First, we investigated the relation between the phase correlations $c(\Delta)$ and 
the NLPE  $\psi (\tau)$ for both time series.
While no obvious trends between linear correlations in the phase maps and the 
nonlinear prediction error could be identified for the DJ data, we found
remarkable anti-correlations for the AAFT  ($-0.68$) and IAAFT  ($-0.94$) 
between  $c(\Delta =1)$ and $\psi (\tau=10 ks)$ for the Mrk 766 data. 
The latter value means a nearly perfect correlation indicating that the NLPE calculated for IAAFT surrogates 
is mostly determined by  Fourier phase correlations for $\Delta = 1$.
We note here that this kind of analysis has the potential to shed more light on the effects of phase correlations on 
the shape of the point distribution in embedding space to ultimately get more insight
into the meaning of Fourier phases for nonlinear time series. For this study, however, it is already sufficient to show that there
{\it are} significant phase correlations which affect the calculation of nonlinear statistics. \\
The results for a standard test on nonlinearities using the NLPE as test 
statistics are summarized in Fig. \ref{figure4}.  We plot the significances $S(\psi)$,
\begin{equation}
S(\psi) = \left| \frac{\psi_{original} - \langle \psi \rangle_{surrogate}}{\sigma_{\psi_{surrogate}}}   \right| \;,
\end{equation} 
as a function of $\tau$. 
The mean $\langle \psi \rangle_{surrogate}$ and the standard deviation 
$\sigma_{\psi_{surrogate}}$ are derived by summing over $N=200$ realizations
of the respective set of surrogates.
Obviously, the propagation of the phase correlations found for the AAFT and IAAFT surrogates 
to the calculation of the NLPE eventually leads to a non-detection of nonlinearities.
For both time series the significances for the AAFT and IAAFT test are well below the 3 $\sigma$  detection criterion, 
whereas the test on dynamic nonlinearities using the remapped time series and FT surrogates 
yields a highly significant detection of nonlinearities with $S(\psi)$ 
reaching eight for Mrk 766 and six for the Dow Jones . 
These results remain stable for a range of delay times.

{\it Conclusions.} Using the examples of an AGN light curve and the Dow Jones day-to-day returns
we demonstrated that both the AAFT and IAAFT algorithm can generate 
surrogate data sets with phase correlations. Thus, these time series have to be considered as nonlinear.
We further showed that those phase correlations propagate into the calculation of nonlinear statistics like 
the NLPE. 
Hence, nonlinearities in time series may remain undetected due to the presence of them in a number of surrogate realizations
against which the original data are compared. 
The wrong outcome of the surrogate test leads then, in turn, to a wrong modeling of the complex underlying system.
In our case the AAFT and IAAFT surrogate test would suggest that the X-ray variability of Mrk 766 
is compatible with a global disk oscillations model, which is clearly ruled out by the FT surrogate test.
Similarly, the detection of weak dynamic nonlinearities (only with FT-surrogates) proves once again that the returns are 
correlated with each other, which disproves one of the basic assumptions of
the Black-Scholes model \cite{Black73, Merton73}, where they are assumed to be random and independent 
from each other. Since the correlations are nonlinear, simple ARMA processes
cannot be used as market models. Finally,  the detected nonlinearities for the 
{\it remapped} data represent signatures that go beyond the 
 so-called stylised facts (see e.g. \cite{Bunde02a}) of the fluctuations of price
indices, that mostly relate to the shape of the probability density derived 
from the original distribution of the return values, e. g. 'fat tails' etc.
Market models must thus not only be able to reproduce these features but must
also account for the intrinsic dynamic nonlinearities detected by means of FT-surrogate test.\\
By analyzing a larger set of AGN light curves as well as other typical 
simulated nonlinear time series like e.g. the Lorenz system 
we convinced ourselves that the presence of phase correlations 
is a generic property of AAFT and IAAFT surrogates -- rather independent 
of the underlying time series under study.\\  
Having outlined the consequences of surrogate tests on model building using the examples 
of an AGN light curve and the Dow Jones day -to-day returns it becomes 
obvious that it may be necessary that previous results obtained with AAFT and IAAFT 
have to be critically reassessed. It may be that weak nonlinearities
remained undetected when AAFT and IAAFT surrogates were used. 
As a consequence wrong physical models may have been found to be compatible 
with the data.\\ 
In general, surrogate generating algorithms aiming at testing both static
and dynamic nonlinearities cannot reproduce both
the power spectrum {\it and } the amplitude distribution while preserving the 
randomness of the Fourier phases so far.
Thus, one has to test separately for static and dynamic nonlinearities.
The latter can be achieved by using FT surrogates for remapped time series as a reliable and truly linear class of surrogates.
Additional comparably easy tests on the Gaussianity of the amplitude distribution can exclude 
the presence of static nonlinearities induced by a nonlinear rescaling.\\ 
{\it Acknowledgments.} This work has made use of observations obtained with {\it XMM-Newton}, an ESA science mission
with instruments and contributions directly funded by ESA member states and the US (NASA).

\bibliography{revisiting_algorithms_for_surros}

\end{document}